\documentclass{article}

\PassOptionsToPackage{numbers, compress}{natbib}

\usepackage[preprint]{neurips_2024}

\usepackage[utf8]{inputenc} %
\usepackage[T1]{fontenc}    %
\usepackage[colorlinks,
    linkcolor=red,citecolor=citecolor,urlcolor=blue]{hyperref}       %
\usepackage{url}            %
\usepackage{booktabs}       %
\usepackage{amsfonts}       %
\usepackage{nicefrac}       %
\usepackage{microtype}      %
\usepackage[table,xcdraw]{xcolor}
\usepackage{algorithm}
\usepackage{graphicx}
\usepackage{threeparttable}
\usepackage{multirow}
\usepackage{amsmath}
\usepackage{bm}
\usepackage{bbm}
\usepackage{amsthm}
\usepackage{harpoon}
\usepackage[]{algpseudocode}
\usepackage{enumitem} %
\usepackage[subrefformat=parens,labelformat=parens]{subfig}

\usepackage{wrapfig}
\usepackage[algo2e, ruled, vlined, linesnumbered, noend]{algorithm2e}
\usepackage{media9}
\usepackage{animate}
\usepackage{soul}

\definecolor{citecolor}{RGB}{34,139,34}
\definecolor{mydarkblue}{rgb}{0,0.08,1}
\definecolor{mydarkgreen}{rgb}{0.02,0.6,0.02}
\definecolor{mydarkred}{rgb}{0.8,0.02,0.02}
\definecolor{mydarkorange}{rgb}{0.40,0.2,0.02}
\definecolor{mypurple}{RGB}{111,0,255}
\definecolor{myred}{rgb}{1.0,0.0,0.0}
\definecolor{mygold}{rgb}{0.75,0.6,0.12}
\definecolor{myblue}{rgb}{0,0.2,0.8}
\definecolor{mydarkgray}{rgb}{0.,0.2,0.2}

\definecolor{lightred}{RGB}{255,235,235}
\definecolor{lightgreen}{RGB}{235,255,235}
\definecolor{lightblue}{RGB}{235,235,255}
\definecolor{lightcyan}{RGB}{235,255,255}
\definecolor{lightmagenta}{RGB}{255,235,255}
\definecolor{lightyellow}{RGB}{255,255,235}

\definecolor{qxkcolor}{RGB}{215,235,255}
\definecolor{softmaxcolor}{RGB}{230,235,255}
\definecolor{probxvcolor}{RGB}{255,255,235}

\definecolor{topkcolor}{RGB}{255,235,235}
\definecolor{zecolor}{RGB}{255,255,235}
\definecolor{dynacolor}{RGB}{235,255,255}

\definecolor{reviewcolor}{RGB}{0,0,200}

\newcommand{\calK}{\mathcal{K}}
\newcommand{\calO}{\mathcal{O}}

\newcommand{\eps}{\vec{\epsilon}}

\renewcommand{\vec}[1]{\boldsymbol{#1}}

\theoremstyle{plain}

\theoremstyle{definition}

\algdef{SE}[DOWHILE]{Do}{doWhile}{\algorithmicdo}[1]{\algorithmicwhile\ #1}%

\graphicspath{{./figs/}}

\newcommand{\name}{\texttt{PIC\textsuperscript{2}O-Sim}\xspace}

\begin{document}

\pagestyle{plain} %

\title{
PIC\textsuperscript{2}O-Sim: A \underline{P}hysics-\underline{I}nspired \underline{C}ausality-Aware Dynamic \underline{C}onvolutional Neural \underline{O}perator for Ultra-Fast Photonic Device FDTD \underline{Sim}ulation
}

\author
{
Pingchuan Ma$^1$,
Haoyu Yang$^2$,
Zhengqi Gao$^3$,\\
\textbf{Duane S. Boning$^3$, Jiaqi Gu$^1$,}
\\
$^1$Arizona State University, $^2$Nvidia, $^3$MIT\\
\small\textit{pingchua@asu.edu},\\
\small\textit{jiaqigu@asu.edu}
}

\maketitle
\begin{abstract}
\label{sec:Abstract}
Optical simulation plays an important role in photonic hardware design flow.
The finite-difference time-domain (FDTD) method is widely adopted to solve time-domain Maxwell equations. 
However, FDTD is known for its prohibitive runtime cost as it iteratively solves Maxwell equations and takes minutes to hours to simulate a single device.
Recently, AI has been applied to realize orders-of-magnitude speedup in partial differential equation (PDE) solving.
However, AI-based FDTD solvers for photonic devices have not been clearly formulated. 
Directly applying off-the-shelf models to predict the optical field dynamics shows unsatisfying fidelity and efficiency since the model primitives are agnostic to the unique physical properties of Maxwell equations and lack algorithmic customization.

In this work, we thoroughly investigate the synergy between neural operator designs and the physical property of Maxwell equations and introduce a physics-inspired AI-based FDTD prediction framework \name.
\name features a causality-aware dynamic convolutional neural operator as its backbone model that honors the space-time causality constraints via careful receptive field configuration and explicitly captures the permittivity-dependent light propagation behavior via an efficient dynamic convolution operator.
Meanwhile, we explore the trade-offs among prediction scalability, fidelity, and efficiency via a multi-stage partitioned time-bundling technique in autoregressive prediction.
Multiple key techniques have been introduced to mitigate iterative error accumulation while maintaining efficiency advantages during autoregressive field prediction.
Extensive evaluations on three challenging photonic device simulation tasks have shown the superiority of our \name method, showing 51.2\% lower roll-out prediction error, 23.5 times fewer parameters than state-of-the-art neural operators, providing 300-600$\times$ higher simulation speed than an open-source FDTD numerical solver. Our code is open sourced at \href{https://github.com/ScopeX-ASU/PIC2O-Sim}{link}.

\end{abstract}

\section{Introduction}
\label{sec:Introduction}

\begin{figure}
    \centering
    \subfloat[]{
    \includegraphics[width=0.242\columnwidth]{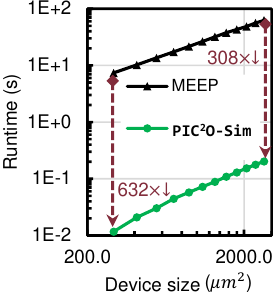}
    \label{fig:MeepRuntime}}
    \subfloat[]{
    \includegraphics[width=0.24\columnwidth]{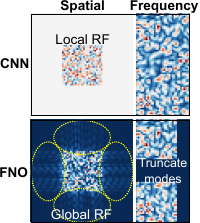}
    \label{fig:LocalGlobalView}}
    \hspace{3pt}
    \subfloat[]{
    \includegraphics[width=0.125\columnwidth]{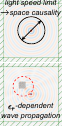}
    \label{fig:MotivationCausality}}
    \hspace{3pt}
    \subfloat[]{
    \includegraphics[width=0.285\columnwidth]{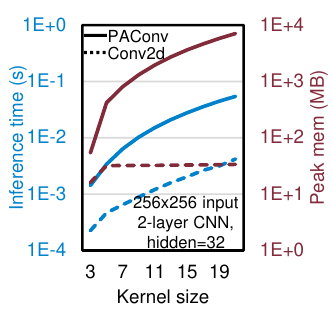}
    \label{fig:ConvMemSpeed}}
    \caption{
    (a) Our ML-based model shows orders-of-magnitude speedup over FDTD solvers.
    (b) Global-view FNO~\cite{li2021fourier} with truncated modes cannot learn local convolution.
    (c) (\emph{Top}) Due to light speed limitation, space causality implies local receptive fields; (\emph{Bottom}) Wave propagation depends on local permittivity distributions.
    (d) Expanding convolutional kernel sizes for long-term prediction is not scalable in runtime and memory, especially for dynamic convolution PAConv~\cite{su2019pixel}.}
    \vspace{-10pt}
\end{figure}

Photonics has shown great potential in high-performance, energy-efficient computing, communication, and sensing due to the fast propagation speed, high bandwidth, and high degree of freedom of photons.
Finite-difference time-domain (FDTD) simulation is a widely adopted numerical method to simulate the spectral response of photonic structures. 
FDTD iteratively solves the time-dependent Maxwell partial differential equations (PDEs) in a discrete mesh to emulate electromagnetic (EM) wave propagation.
However, FDTD simulation is considerably time-consuming as it updates the EM field distribution repeatedly on a mesh with high space-time resolutions due to convergence considerations.
The computational complexity of 2D FDTD is $\calO(N^2T)$, where $N$ and $T$ are spatial grid dimensions and simulation timesteps, respectively.
For example, even simulating a small 2D photonic device of size $28 \mu m \times 10.5 \mu m$ over a 1.3 ps timespan costs around 1 min on powerful CPUs, as shown in Fig.~\ref{fig:MeepRuntime}.
Simulating large-scale devices/circuits with fine-grained structures and resonant responses requires hours of computing time, which significantly slows down the design process and makes it intractably expensive for iterative simulation-in-the-loop inverse optimization.
It is in high demand to develop an ultra-fast surrogate simulation model to obtain rapid early-stage feedback during design.

There is a promising trend to employ AI to speed up PDE solving for physical simulation.
Physics-informed neural networks (PINNs)~\cite{NN_PINN_2019} and neural operators (NOs) have been demonstrated to learn the nonlinear high-dimensional functional mapping from PDE observations to the PDE solution, showing orders-of-magnitude faster speed than numerical solvers.
AI-based surrogate models have been demonstrated in a variety of scientific applications, such as flow dynamics/field simulation~\cite{li2021fourier, tran2023factorized,JMLR:v24:21-1524, ML_Arxiv2022_Ashiqur,ovadia2023realtime, taccari2023understanding, yang2023denoising}, weather forecasting~\cite{NN_PASC23_Kurth}, and hardware simulation and inverse design~\cite{NP_Nature2021_Chen, NP_APL2022_Lim,NP_NatureCompSci2022_Tang, gu2022neurolight,NP_ACSPhotonics2023_Augenstein}.
In the field of AI for optics, 
prior work has introduced physics-augmented or data-driven neural networks~\cite{NP_APL2022_Lim,NP_Nature2021_Chen,gu2022neurolight} on finite-difference frequency-domain (FDFD) optical simulation.
Some work has explored time-domain field prediction to replace FDTD solvers with PINNs~\cite{NN_IEEEJAP2020_Noakoasteen,NN_WMCS2020_Zhang} and physics-driven recurrent neural networks (RNNs)~\cite{NP_SciAdv2019_Hughes} that incorporate the iterative FDTD updating rules into the model architecture to directly model the light propagation.

It remains under-explored that AI can be used to solve the time-domain Maxwell equations and generate optical field dynamics directly as a video.
In this work, we mainly answer several key questions for AI-based optical FDTD tasks.
(1) \textbf{What neural operator architecture honors the physics constraints of time-domain Maxwell equations?}~
State-of-the-art neural operators designed for flow dynamics or EM wave simulation generally employ global-view operators, e.g., multi-scale convolution~\cite{zhang2024sinenet, unet} and Fourier-domain operators~\cite{li2021fourier}, to capture the long-distance correlation in the solving domain, shown in Fig.~\ref{fig:LocalGlobalView}.
However, this may violate the \underline{space-time causality constraints} of Maxwell equations, where the light speed enforces the maximum distance information can propagation, shown in Fig.~\ref{fig:MotivationCausality} (\emph{Top}).
Moreover, Fourier operators with truncated frequency modes lack the ability to represent a local-view operation, which makes convolution with a restricted receptive field a better candidate.
(2) \textbf{How to effectively represent PDE variables into the model architecture?}
Most prior work feeds all the PDE variables as a multi-channel tensor into the neural operator and ignores each variable's physical property.
Specifically, Maxwell equations imply a \underline{permittivity-dependent wave propagation} behavior, shown in Fig.~\ref{fig:MotivationCausality}(\emph{Bottom}), that can hardly be captured by static matrix multiplication or convolution.
(3) \textbf{What are the trade-offs among scalability, long-term prediction fidelity, and speedup?}~
Since FDTD requires generating a video of field dynamics, it is unclear how to formulate this video generation task in a \underline{scalable, efficient, and high-fidelity} fashion.
An intuitive understanding is that it is easy to predict the single frame in the next timestep with high fidelity but loses speed benefit due to a large number of iterations and suffers from large roll-out errors due to error accumulation over time.
Predicting the entire video in one shot eliminates error accumulation but significantly increases the learning difficulty and is ultimately not scalable to handle a long simulation timespan as the computing and memory cost quadratically increase with larger receptive fields (kernel sizes) shown in Fig.~\ref{fig:ConvMemSpeed}. 

Based on the above observations on the unique property of optical FDTD simulation, we answer the above three key unresolved questions and introduce a physics-inspired data-driven ML-based photonic FDTD simulation framework, dubbed \name.
Motivated by light propagation's space-time locality and permittivity dependency, we find an analogy between optical FDTD simulation and dynamic convolution.
Our framework \name features a dynamic convolutional neural operator model to predict high-quality light fields in a time-bundled way.
To generate long timespan light field dynamics, we adopt an autoregressive, multi-iteration method to maintain scalability and speed advantages while minimizing the long-term prediction error.

The main contribution of this work is three folds:
\begin{itemize}[leftmargin=*]
\setlength{\itemindent}{0.5em}
\item We deeply investigate the unique properties and unresolved challenges in AI-based time-domain Maxwell equation solving and propose a physics-inspired causality-constrained photonic device FDTD simulation framework \name with balanced scalability, prediction fidelity, and speedup.
\item We point out the analogy between the FDTD and dynamic convolutions and design a convolutional neural operator model with causality-aware receptive field designs and dilated position-adaptive dynamic convolutions\cite{su2019pixel,xu2021paconv,pmlr-v189-hachiya23a} for permittivity-aware light propagation prediction.
\item We propose an autoregressive framework for scalable long-term video generation and mitigate the temporal error accumulation via cross-iteration error correction techniques. 
\item Extensive evaluation has shown our superior prediction fidelity and efficiency on three types of complicated photonic devices, showing \textbf{51.2\%} lower prediction errors, \textbf{23.5$\times$} higher parameter efficiency, and \textbf{308$\times$-632$\times$} faster than an open-source FDTD solvers Meep.

\end{itemize}

\section{Background: AI for PDE Solving}
\label{sec:Background}
\vspace{-5pt}
Recently, scientific machine learning algorithms have been widely explored to help solve fundamental PDE problems with orders-of-magnitude faster speed.
PINNs and data-driven neural operators represent two branches of research where physics is either added as a hard constraint or ignored to remove domain knowledge requirements.
In the field of AI for optics, physics-informed models, e.g., WaveTorch~\cite{hughesWavePhysicsAnalog2019}, directly embed the PDE updating rules in the recurrent neural network (RNN) cells to leverage the GPU-accelerated inference engine for faster iteration.
With a small enough spatiotemporal resolution, these methods have a theoretical guarantee on the solved fields, while their speedup is rather limited due to a large number of iterations.
Also, oversimplified equations in the RNN cells make it hard to match the golden results from commercial tools.

Physics-augmented models, e.g., MaxwellNet~\cite{NP_APL2022_Lim}, WaveYNet~\cite{NP_Nature2021_Chen}, adopted a standard U-Net structure and incorporated Maxwell residual loss in the training objective to learn an optical field that honors physical constraints.
Recently, SineNet~\cite{zhang2024sinenet} was proposed to mitigate the temporal misalignment caused by the skip connection between multi-scale features in U-Net by cascading multiple U-Net and hence reducing the temporal misalignment.

However, the synergy between model architecture and the underlying physical constraints of Maxwell equations remains under-explored.
SoTA neural operators might not suit the optical FDTD due to the unique properties of Maxwell equations.
Besides, prior work often focuses on single-iteration prediction tasks without handling the error accumulation effects with autoregressive prediction.

\section{Proposed \name Framework}
\label{sec:Method}
\subsection{Understanding the FDTD simulation for light propagation in photonic devices}
\label{sec:Formulation}
First, we formulate the FDTD method for photonic device simulation.
FDTD starts by injecting an eigenmode light source into the device and simulating the light propagation via sequential time-marching.
To obtain the response at multiple wavelengths in one shot, The incident source typically has a Gaussian-shaped envelope centered at frequency $f_c$ with a frequency width of $f_w$, thus carrying a wide range of wavelengths for broadband simulation.
FDTD method discretizes the time-domain Maxwell equation and iteratively update of the electric fields. For the detail of the electric field updating rule, please refer to Appendix~\ref{sec:update_rule}

\noindent\textbf{Considerable computational complexity of FDTD}.~FDTD is time-consuming as its convergence depends on fine-grained space-time resolution ($\Delta_t,\Delta_x,\Delta_y$) to capture the light-speed signal propagation accurately, e.g., a typical timestep is $\Delta_t=0.167 fs$, and a space resolution is around 1/15 of the wavelength, leading to high computational complexity of $\calO(\frac{N_xN_yT}{\Delta_x\Delta_y \Delta_t})$, where $N_x,N_y$ are solving domain dimension and $T$ is the simulation timespan.
Usually, to improve convergence, electric and magnetic fields will be alternatively updated on an interleaved 2D Yee's grid,
which further increases the computation cost by 4 times.
Hence, a fast prediction method that can skip tens of thousands of FDTD time-marching steps and directly reconstruct the spatio-temporal field dynamics will significantly speed up time-domain photonic device simulation.

\noindent\textbf{Causality-constrained space-time locality}.~
Near-future light field prediction tasks constrain wave%
\begin{wrapfigure}[15]{rH}{0.53\textwidth}
\begin{minipage}{0.53\textwidth}
    \centering
    \vspace{-13pt}
    \includegraphics[width=0.99\columnwidth]{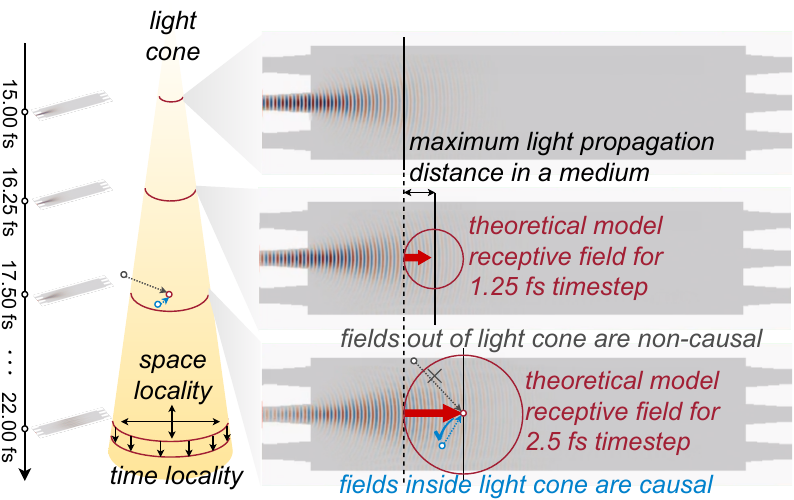}
    \caption{Illustration of the causality-constrained space-time locality and the theoretical receptive field.}
    \label{fig:VisualizeReceptiveField}
\end{minipage}
\end{wrapfigure}
propagation in a local spatial region due to limited light speed, as indicated by the light cone in Fig.~\ref{fig:VisualizeReceptiveField}.
This indicates that any fields outside the light cone will be non-causal to the center field, which implies a confined theoretical spatial receptive field (RF) of any prediction model.
A model with an overly small RF lacks the information to reconstruct the center field.
Similarly, a model with an overly large RF or even global views beyond the light cone can potentially learn non-physical mapping as it mixes irrelevant, non-causal information.
Besides spatial locality limited by light speed, the light field shows temporal locality.
Based on Maxwell equations, the field distribution at timestep $t$ solely depends on the electromagnetic wave in the previous timestep $t-1$, which indicates that it is theoretically unnecessary to capture a long context in the model design as preferred in other sequence/time-series modeling tasks.
Therefore, it is important to design a model with a carefully selected space-time receptive field that honors causality. 

\noindent\textbf{Permittivity-dependent light propagation}.~
The aforementioned causality-constrained space-time locality indicates that it is suitable to employ a convolution-based neural operator with a carefully selected spatial receptive field to propagate the light waves from a causal neighboring region to%

\begin{wrapfigure}[19]{rH}{0.55\textwidth}
\begin{minipage}{0.55\textwidth}
    \centering
    \vspace{-15pt}
    \includegraphics[width=\columnwidth]{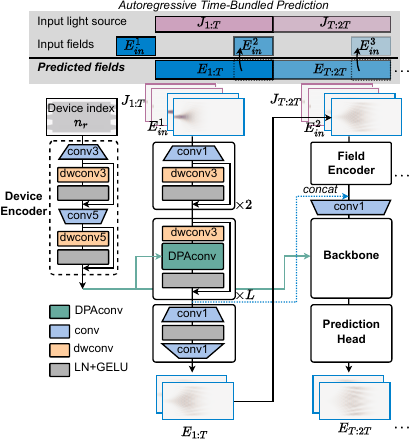}
    \caption{\name framework overview.
    }
    \label{fig:ModelArch}
\end{minipage}
\end{wrapfigure}
the center location.
However, static convolution operations may not be sufficient to model the light-matter interaction in the photonic device as the local wave propagation behavior at coordinate [$m,n$] is a function of material permittivity $\epsilon[m,n]$ at that specific location, shown in Fig.~\ref{fig:MotivationCausality}. 
Such a property implies the convolutional filter that emulates the wave propagation mechanism should contain dynamic values based on the local material permittivity, which inspires us to propose a dynamic permittivity encoding in the convolutional filters.

\subsection{Proposed \name framework}

\subsubsection{Framework overview}
Figure~\ref{fig:ModelArch} illustrates our autoregressive time-bundled \name framework.
To potentially handle prediction over a long time horizon, \name autoregressively predicts the future optical field dynamics based on previous fields $E_{in}$ and corresponding light source $J$.
Time-bundling is employed to predict multiple timesteps/frames of fields at each iteration.

\name formulates a single-iteration FDTD prediction task as a functional mapping from given initial condition $\mathcal{A} \in \mathbb{R}^{\Omega \times d_a}$, including previous light fields, light source, and device permittivity distribution, to the electrical field solution $\mathcal{U} \in \mathbb{R}^{\Omega \times d_u}$. 
$\Omega$ is the 2-D field domain with a size of $[N_x, N_y]$. 
\name takes $E_{in}$ from previous $T_{in}$ timesteps and $J_{1:T}$ in the future $T$ timesteps as input and passes through a convolutional neural operator $\Psi_{\theta}$ to predict the future $T$-step light fields $E_{1:T}$.
The neural operator $\Psi_{\theta}$ consists of a field encoder, a dynamic convolutional backbone, and a prediction head.
Permittivities/refractive indices of the device are explicitly encoded by a dedicated device encoder and fed into all dynamic position-adaptive convolutional layers in the model backbone to guide light propagation.

\subsubsection{Resolution-preserved shift-invariant domain discretization}
To deal with various sizes of the input device, previous work~\cite{gu2022neurolight} adopted a scale-adaptive domain%
\begin{wrapfigure}[15]{rH}{0.28\textwidth}
\begin{minipage}{0.28\textwidth}
    \centering
    \vspace{-8pt}
    \includegraphics[width=\columnwidth]{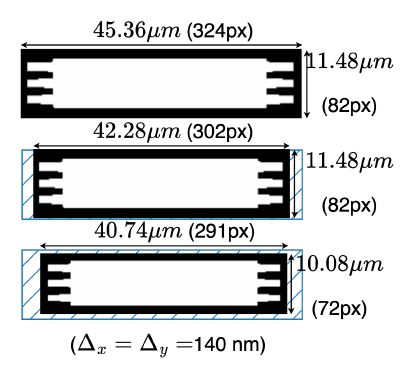}
    \caption{Devices in a mini-batch are replicate-padded to the same size with a fixed spatial resolution.}
    \label{fig:DomainDiscretization}
\end{minipage}
\end{wrapfigure}
discretization that scales devices of all physical sizes to the same image size since Fourier neural operators formulate the function mapping in a fixed domain $\Omega$. 
In contrast, our \name adopts preserves the resolution ($\Delta_x,\Delta_y$) with padding if necessary to maintain shift-invariance.
This design choice brings two major advantages:
(1) \textbf{Shift-invariant is more scalable and generalizable than resolution-invariant discretization in our problem}.~The prediction model is trained on ground-truth fields simulated with high enough resolutions to guarantee FDTD accuracy, i.e., a wavelength contains 15-20 pixels.
Predicting fields at higher resolution does not pragmatically bring benefits.
Furthermore, a clear issue of downsizing is its inability to handle unseen large devices, where a large downsampling factor will cause severe information loss.
The shift-invariant property of convolution, instead, allows the model to predict light propagation in an arbitrarily large domain without downsampling-induced loss as long as the spatial resolution stays the same.
(2) \textbf{Training efficiency benefit}.~
To avoid downsampling-induced information loss, the domain-adaptive method tends to scale all devices to a large image size, which causes high costs during training and inference.
In contrast, \name scales all devices to the same pixel resolution, e.g., $\Delta_x=\Delta_y=140 nm$ and pads devices to the maximum image size only in this mini-batch for parallel batched processing as shown in Fig.~\ref{fig:DomainDiscretization}.
In this way, we can avoid information loss due to the downsampling of large devices and improve speeds on small devices.

\subsubsection{Model input/output definition: permittivity, input fields, and sources}
At one iteration, the model takes the permittivity $\epsilon_r$ and $T_{in}$-frame input fields $E_{in}\in\mathbb{R}^{T_{in}\times H\times W}$%
\begin{wrapfigure}[12]{rH}{0.22\textwidth}
\begin{minipage}{0.22\textwidth}
    \centering
    \vspace{-10pt}
    \includegraphics[width=\columnwidth]{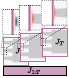}
    \caption{Light source representation.}
    \label{fig:MaskedInjectLight}
\end{minipage}
\end{wrapfigure}
before the target prediction timestep, and $T$-frame light sources $J_{1:T}$ as the PDE variables.
For 2D simulation, the line-shaped Gaussian eigenmode source $J_{1:T}$ is injected at the port center.
The existence of a source in the system is the fundamental difference and also the challenging part compared to other source-free PDE systems.
The right-hand side of the Maxwell equation is not zero but a time-varying function $J(t)$, such that each frame of field is potentially impacted by all previous injected light sources.
We formulate the variable light source within the prediction time horizon as a $T$-frame video, where all fields are masked to zero except the line-shaped region of light source at the input port.
In this way, the FDTD prediction task is translated to a masked video restoration task, given that previous frames and the video patches at the source location are unmasked hints.

\subsection{Efficient physics-inspired dynamic convolutional neural operator architecture}
\label{dynamic_convolution}
Based on the space-time causality in Section~\ref{sec:Formulation}, our \name model $\Psi_{\theta}$ is built with local-view convolutions to restrict the receptive field.
We introduce the detailed architecture as follows. 

\noindent\textbf{Convolutional field encoder}.~
\name starts with a convolutional encoder to project the previous light fields and incident light sources to a $D$-dimensional latent space: $a_E^{\dagger}(\boldsymbol{r}) \rightarrow v_0(\boldsymbol{r}), \forall \boldsymbol{r} \in \Omega$, where $a_E^{\dagger}=\{E_{in};J_{1:T}\} \in \mathbb{R}^{(T_{in} + T) \times M \times N}$ and $v_0(\boldsymbol{r}) \in \mathbb{R}^{D \times M \times N}$.
The encoder has two blocks, each containing a point-wise convolution followed by a residual block of 3$\times$3 depthwise convolution, layer normalization, and the GELU activation function.

\noindent\textbf{Causality-constrained permittivity-aware convolutional backbone}.~
The backbone of \name consists of $L$-layer residual blocks, each including a depthwise convolution, dilated position-adaptive convolution (DPAConv), layer normalization, and GELU activation function. 
A dedicated convolutional device encoder shown in Fig.~\ref{fig:ModelArch} takes the inverse of permittivity map $1/\epsilon_r(r)$ as input and extracts a shared local geometry information $\boldsymbol{f}(\epsilon_r^{-1})$ for all $L$ DPAConv layers in the backbone.

\begin{wrapfigure}[16]{rH}{0.55\textwidth}
\begin{minipage}{0.55\textwidth}
    \centering
    \vspace{-12pt}
    \includegraphics[width=\columnwidth]{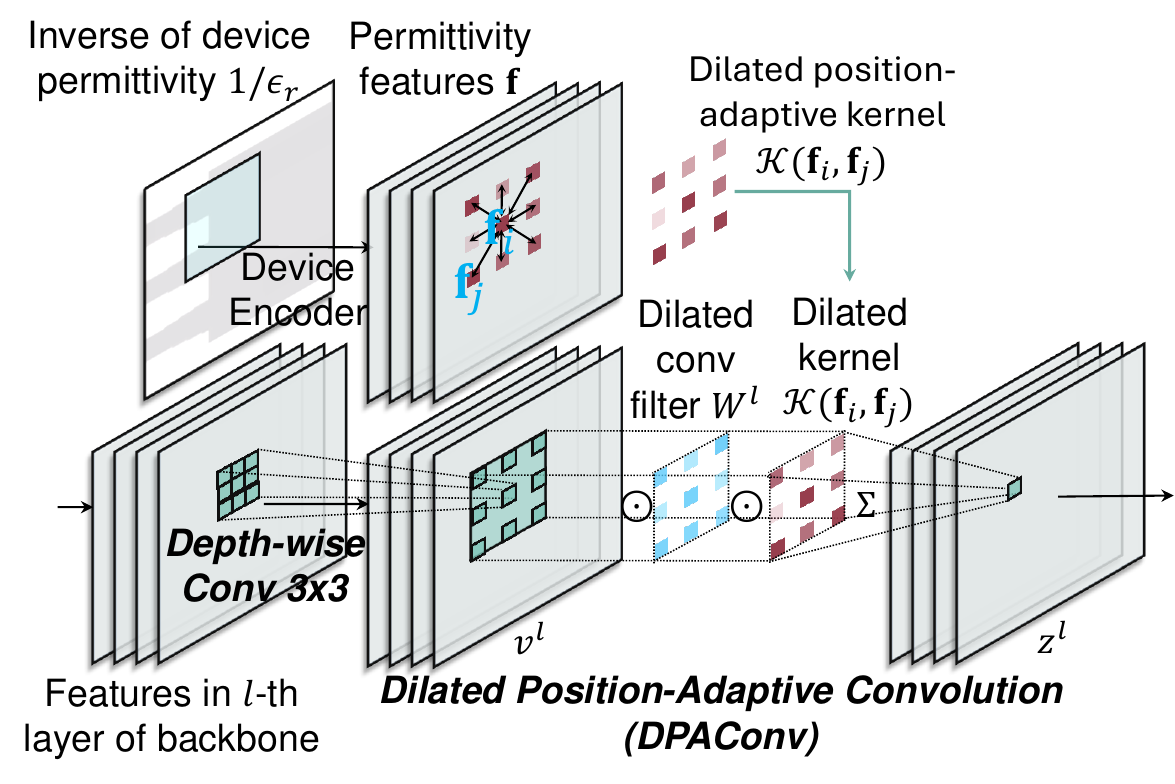}
    \caption{Illustration of the proposed dilated position-adaptive convolution (DPAConv).}
    \label{fig:DPAConv}
\end{minipage}
\end{wrapfigure}

Figure~\ref{fig:DPAConv} illustrates a $K \times K$ DPAConv module with dynamic permittivity-adaptive kernels. 
Each DPAConv operation within a size-$K$ window $\Omega(i)$ at pixel position $i$ is formulated as $z^{l}(i)=\sum_{j \in \Omega(i)} \big(\calK\left(\mathbf{f}_i, \mathbf{f}_j\right)(\epsilon_r) \mathbf{W}^T(j)\big) \cdot v^l(j)$.
Inspired by dynamic convolution (PAConv)~\cite{su2019pixel}, the convolutional filter weights applied to a sliding%
window on the feature map is the Hadamard product of a statically-learned convolutional filter $W \in \mathbb{R}^{D\times D \times K \times K}$ and a dynamic permittivity-adaptive kernel $\mathcal{K}(\mathbf{f}_i, \mathbf{f}_j)(\epsilon_r) \in \mathbb{R}$, shown in Fig.~\ref{fig:DPAConv}.
The dynamic kernel $\calK$ projects the permittivity features $\mathbf{f}$ into high-dimensional space via a Gaussian kernel function $\calK\left(\mathbf{f}_i, \mathbf{f}_j\right)(\epsilon_r)=\exp \left(-\frac{1}{2}\left(\mathbf{f}_i-\mathbf{f}_j\right)^{\top}\left(\mathbf{f}_i-\mathbf{f}_j\right)\right)$, which helps the model understand light-matter interaction and learn how light wave propagates dynamically through a path with heterogeneous material permittivities.

This dynamic convolution shows strong modeling capability to capture wave propagation principles. 
However, standard PAConv~\cite{su2019pixel} causes a practical challenge with considerable memory and runtime costs, especially during training.
As the prediction frames $T$ increase, the required receptive field and convolutional kernel size increase linearly.
When a large-kernel convolution requires dynamic position-specific kernels, the memory cost is bottlenecked by the largest%
\begin{wrapfigure}[8]{rH}{0.4\textwidth}
\begin{minipage}{0.4\textwidth}
    \centering
    \vspace{-10pt}
    \includegraphics[width=\columnwidth]{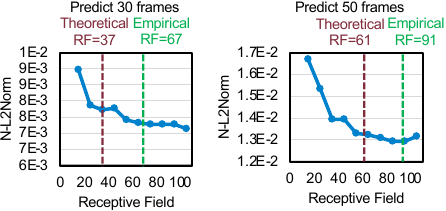}
    \caption{30 pixels larger receptive field could provide optimum fidelity}
    \label{fig:ReceptiveField_small}
\end{minipage}
\end{wrapfigure}
intermediate tensor $\calK\odot W\in\mathbb{R}^{D\times H\times W \times K\times K}$, e.g., when $D=96$, $H=W=256$, $K=21$, a single tensor takes >10 GB. 
To reduce the computation and memory burden, we modify it as a dilated position-adaptive convolution with an additional depthwise convolution ahead of it to aggregate local features and avoid information loss from dilation.
Kernel size $K$ is the key design parameter for DPAConv that determines the receptive fields of the model. As Figure~\ref{fig:ReceptiveField_small} shows, insufficient receptive field degrades the performance intensively, we empirically suggest using \textbf{a receptive field that is 30 pixels larger than the theoretical value}. 
For detailed kernel size selection, please see Appendix~\ref{sec:HyperparameterPicking}

\noindent\textbf{Prediction head}.~ 
At the end of the model, we simply use two point-wise convolutions with layer normalization and GELU in between to project it back to the required prediction frames.

\subsection{Autoregressive prediction with multi-stage partitioned time-bundling}
\vspace{-3pt}
\label{sec:Multistage}
Time-bundling significantly speeds up the prediction as it generates multiple frames at one shot and reduces iterations during autoregression~\cite{brandstetterMessagePassingNeural2023}.
However, we claim \textbf{bundling too many timesteps is harmful to scalability and prediction fidelity}.
(1) \underline{Scalability}:
Given that the required RF linearly expands when predicting more output frames, it has \emph{quadratic parameter count}, \emph{quadratic computation cost}, and \emph{higher optimization difficulty} with larger convolutional kernels.
Moreover, predicting more frames requires more model capacity; the hidden dimension $D$ also needs to increase accordingly.
Hence, it is \emph{not scalable to time-bundle too many frames} for each iteration of model inference.
(2) \underline{Prediction Fidelity}:
Time-bundling essentially shares the learned encoder and backbone between different timesteps for efficiency consideration and inevitably correlates different output frames.
Since the correlation between frames gets reduced with a longer timespan, such sharing can deteriorate the overall prediction fidelity. 
Besides, bundling too many timesteps in a one-shot prediction also \emph{breaks the temporal causality} in light waves, as output field $E[t]$ strictly should not see any information from future sources $J_{t+1:T}$.
Hence, reducing bundled timesteps by partitioning the frames into multiple stages can largely relax the scalability and causality issue while still benefiting from the speed advantages in parallel time-bundled prediction.

Figure~\ref{fig:ModelArch} illustrates our multi-stage time-bundling in autoregressive prediction.
Denote the total predicted fields $E_{1:NT}$ have $NT$ frames, and we partition it into $N$ stages.
In standard autoregressive prediction, the model $\Psi_{\theta}$ will be trained on a single iteration and will recurrently generate future fields based on previous-stage field predictions.
This regularizes the model to learn consistent mapping functions across timesteps.
However, this assumes an ideal single-stage prediction fidelity.
As the first stage prediction $E_{1:T}$ deviates from the ground-truth $E^*_{1:T}$, the autoregression will have \emph{accumulated roll-out error} since the inputs to the model in later iterations will have \emph{distribution shifts} due to prediction errors from previous stages.
To solve this issue, we propose two methods to mitigate the temporally accumulated error.

\noindent\textbf{Stage-dedicated prediction models to mitigate distribution shift}.~
We use $N$ independent submodels $\Psi_{\theta_1}, \cdots, \Psi_{\theta_N}$ in the partitioned $N$ stages to mitigate the distribution shift issue.
Stage 2 to stage $N$ will learn how to align the non-ideal input fields from their previous stage to the ground-truth fields.
In a later experiment section, we will show the advantages of independent models in reducing prediction errors while maintaining high parameters and runtime efficiency.

\noindent\textbf{Cross-stage hidden state propagation to facilitate error mitigation}.~
Theoretically, the time locality of the Maxwell equation implies that the light fields depend only on the near-past fields and future sources.
However, the non-ideal field predictions from $\Psi_{\theta_i}$ do not carry enough information for $\Psi_{\theta_{i+1}}$ to compensate all errors from the $i$-th stage.
Inspired by the State-Space Model, we add an extra information path by propagating the encoded hidden states after the backbone of the $i$-th stage to the input of the next-stage backbone.
An additional point-wise convolution is used as a lightweight adaptor to compress the concatenated hidden states back to hidden dimension $D$.

\noindent\textbf{Light field normalization}.~Normalization is critical for model convergence and generalization.
Empirically, we find that using the maximum field intensity to normalize the input fields and source provides the best roll-out error. 
For a detailed ablation study, please refer to Appendix~\ref{sec:HyperparameterPicking}

\section{Result}
\label{sec:Results}
\vspace{-8pt}
\subsection{Experimental Setup}
\vspace{-8pt}
\noindent\textbf{Benchmarks}.~
We evaluate different methods on three representative and challenging photonic device types, including tunable multi-mode interference (MMI) with complicated interference patterns, micro-ring resonator (MRR) with sensitive coupling and resonance effects, and Metaline with highly discrete permittivity distributions and fine-grained structures.
Those practical devices post significant challenges and haven't been evaluated in the literature.
We use the open-source FDTD software package MEEP~\cite{NP_MEEP_2010} to generate the simulation videos.
All videos are resized to have a spatial resolution of ($\Delta_x=\Delta_y=140 nm$, $\Delta_t=1$ fs).
Details are in the Appendix~\ref{tab:AppendixDataset}.

\noindent\textbf{Training settings and evaluation metrics}.~
Since all frames have the same importance in FDTD, we use averaged per-frame normalized L2-Norm as the training loss function and also evaluation metric, i.e., $\texttt{N-L2Norm}=\frac{1}{T}\sum_{t=1}^{T}\|\Psi_{\theta}(E_{in},\epsilon_r,J_{1:T})[t]-E^{\ast}[t]\|_2/\|E^{\ast}[t]\|_2$.
We use frames per second (FPS) to evaluate the prediction speed.
Detailed training settings can be found in Appendix~\ref{sec:AppendixTrainingSettings}.

\subsection{Main Result}
\vspace{-5pt}
We compare 7 models in Table~\ref{tab:MainResults}, including (1) \emph{global-view Fourier-domain neural operators}: FNO~\cite{li2021fourier} and its factorized variant F-FNO~\cite{tran2023factorized}, the SoTA optical FDFD NN surrogate NeurOLight~\cite{gu2022neurolight}, a Koopman neural operator (KNO) that models the time marching in the linear Koopman space~\cite{xiong2023koopman}; and \emph{local-view convolution-based neural operators}: a 16-layer SimpleCNN with static 2D convolution, SineNet~\cite{zhang2024sinenet} with a cascaded multi-stage UNet structure for temporal modeling, and our proposed dynamic convolutional neural operator \name. 
Note that for a fair comparison, the full mode is used in all Fourier-domain neural operators to enable them to learn local spatial operations.
The video length spans 160 fs, i.e., 160 frames, with 10 frames of past input fields ($T_{in}$=10) as initial conditions.
Detailed configurations are in Appendix~\ref{sec:AppendixModelDetails}.

Compared to these baselines, on average, our \name achieved 51.2\% less normalized L2-norm error with 95.7\% fewer parameters

\begin{table}[]
\centering
\caption{Compare different models on three benchmarks in terms of parameter count, inference speed (FPS), training, and test error (N-L2Norm).
The predicted light fields have 160 frames.
}
\label{tab:MainResults}
\resizebox{0.85\columnwidth}{!}{
\begin{tabular}{cc|cccc}
\hline
Dataset                                                               & Model                        & \#Params $\downarrow$                    & FPS $\uparrow$                          & Train error $\downarrow$                  & Test error $\downarrow$                   \\ \hline
                                                                      & FNO~\cite{li2021fourier}                          & 340M                         & 8147                          & 0.035                         & 0.122                           \\
                                                                      & F-FNO~\cite{tran2023factorized}                        & 4.5M                         & 4359                          & 0.039                         & 0.070                         \\
                                                                      & KNO~\cite{xiong2023koopman}                          & 171.8M                       & 251                           & 0.188                         & 0.193                         \\
                                                                      & NeurOLight~\cite{gu2022neurolight}                   & 2.2M                         & 8180                          & 0.157                         & 0.140                          \\
                                                                      & SimpleCNN                    & 3.8M                         & 17524                         & 0.066                         & 0.075                         \\
                                                                      & SineNet~\cite{zhang2024sinenet}                      & 38M                          & 3414                          & 0.071                         & 0.085                         \\
\multirow{-8}{*}{\begin{tabular}[c]{@{}c@{}}MMI\\ \includegraphics[width=3cm]{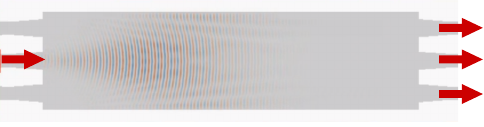}\end{tabular}}& \cellcolor[HTML]{EFEFEF}\name & \cellcolor[HTML]{EFEFEF}2.4M & \cellcolor[HTML]{EFEFEF}15701 & \cellcolor[HTML]{EFEFEF}0.042 & \cellcolor[HTML]{EFEFEF}0.052 \\ \hline
                                                                      & FNO~\cite{li2021fourier}                          & 340M                         & 8147                          & 0.033                         & 0.423                         \\
                                                                      & F-FNO~\cite{tran2023factorized}                        & 4.5M                         & 4376                          & 0.028                         & 0.138                         \\
                                                                      & KNO~\cite{xiong2023koopman}                          & 171.8M                       & 1252                          & 0.138                         & 0.179                         \\
                                                                      & NeurOLight~\cite{gu2022neurolight}                   & 2.2M                         & 8190                          & 0.102                         & 0.151                         \\
                                                                      & SimpleCNN                    & 7.3M                         & 7646                          & 0.038                         & 0.088                         \\
                                                                      & SineNet~\cite{zhang2024sinenet}                      & 38M                          & 3282                          & 0.044                         & 0.109                         \\
\multirow{-7}{*}{\begin{tabular}[c]{@{}c@{}}MRR\\ \includegraphics[width=2cm]{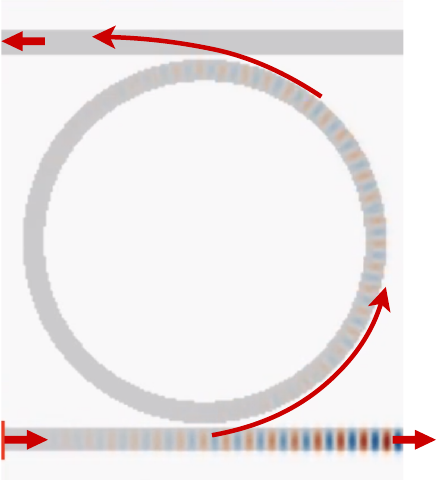}\end{tabular}}     & \cellcolor[HTML]{EFEFEF}\name & \cellcolor[HTML]{EFEFEF}4.4M & \cellcolor[HTML]{EFEFEF}1906  & \cellcolor[HTML]{EFEFEF}0.025 & \cellcolor[HTML]{EFEFEF}0.085 \\ \hline
                                                                      & FNO~\cite{li2021fourier}                          & 146.4M                         & 20047                         & 0.062                         & 0.173                         \\
                                                                      & F-FNO~\cite{tran2023factorized}                        & 3.3M                         & 9713                          & 0.053                         & 0.089                         \\
                                                                      & KNO~\cite{xiong2023koopman}                          & 74.6M                        & 2889                          & 0.278                         & 0.268                         \\
                                                                      & NeurOLight~\cite{gu2022neurolight}                   & 1.6M                         & 18413                         & 0.213                         & 0.185                         \\
                                                                      & SimpleCNN                    & 3.8M                         & 26920                         & 0.112                         & 0.117                         \\
                                                                      & SineNet~\cite{zhang2024sinenet}                      & 30M                          & 4484                          & 0.114                         & 0.122                         \\
\multirow{-7}{*}{\begin{tabular}[c]{@{}c@{}}Metaline\\ \includegraphics[width=3cm]{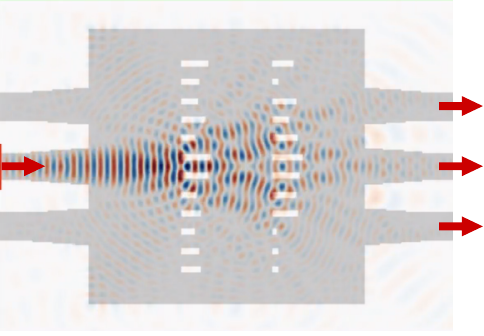}\end{tabular}}       &\cellcolor[HTML]{EFEFEF}\name & \cellcolor[HTML]{EFEFEF}2.4M & \cellcolor[HTML]{EFEFEF}7348  & \cellcolor[HTML]{EFEFEF}0.077 & \cellcolor[HTML]{EFEFEF}0.086 \\ \hline
\multicolumn{2}{c|}{Avg Improv.}                                                                     & 95.74\%                         & 4.4\%                        & 53.33\%                       & 51.16\%                       \\ \hline
\end{tabular}
}
\vspace{-10pt}
\end{table}

\noindent\textbf{Result in tunable MMI}.~
On tunable MMI, \name achieves the most accurate prediction result with only 2.4M parameters, less than 1\% of full-mode FNO. 
Without the inductive bias of the local spatial receptive field, FNO requires full frequency modes to learn a causality-aware local window. 
This comes at the cost of hundreds of millions of parameters and optimization challenges in large kernel learning.
SineNet, although it is a U-Net-based model, and the receptive field is too large due to its multi-scale feature fusion that introduces too many irrelevant features, shows 60\% more error compared to \name. 
Compared to the Fourier-domain models and U-Net-based model, SimpleCNN and \name show better performance, and we attribute the superiority to the causality inductive bias. 
SimpleCNN, due to the content-agnostic static kernels, shows slightly worse fidelity than \name.
This is due to the homogeneity of permittivity in the major part of MMI.

\noindent\textbf{Result in MRR}.~
Compared with MMI, MRR is more difficult as it includes light coupling, feedback loop, and resonance. 
Considering that the $\epsilon_r$ in MMR is smaller than that in MMI, which means that the speed of light is faster, we changed the receptive field according to CNN-based networks. 
Still, an approximately 50\% performance gap shows between the Fourier-domain models, the U-Net-based models, and the CNN-based models. 
Among all the CNN-based models, our \name achieved the lowest error with one of the fewest parameters

\noindent\textbf{Result in Metaline}.~
Compared to MMI, Metaline has highly discrete permittivity distributions with strong scattering effects, which require dynamic convolution to capture the wave propagation behavior accurately.
As a result, we indeed observe a high prediction error from static convolutional networks.

On average, our \name achieved 51.2\% less test error with 95.7\% fewer parameters, showing the advantages of our dedicated device encoder and DPAConv-based model backbone. 

\subsection{Ablation Study and Discussion}

\noindent\textbf{Input field frames $T_{in}$ and dilation rate $s$}.~As a key hyperparameter, we select 10 frames of input fields $E_{in}$, and we choose a dilation rate of 4 to balance fidelity, speed, and parameter efficiency. 
For details, please refer to Appendix~\ref{sec:HyperparameterPicking}.

\noindent\textbf{Multi-stage partitioning in time-bundled prediction}.~
As shown in Table~\ref{tab:partition}, we explored different numbers of multi-stage partitions when the model performs time-bundled prediction on a total of 160 frames of fields. 
Different stages have independent $\Psi_{\theta}$.
A single stage performs a one-shot prediction of 160 frames.
However, the model capacity is not enough to bundle so many frames of fields with high fidelity.
Also, it is not scalable to handle long-range predictions. %

\begin{wrapfigure}[11]{rH}{0.63\textwidth}
\begin{minipage}{0.63\textwidth}
\vspace{-2pt}
\captionof{table}{Compare different task partitioning when predicting 160 frames.
Kernel sizes (\emph{KS}) are adjusted to match the suitable receptive field for the predicted frames ($T$) per step.
Partitioning into 2 stages gives the best results.}
\resizebox{\columnwidth}{!}{
\begin{tabular}{ccc|cccc}
\toprule
                     \#stages & $T$ per stage & KS & \#Params $\downarrow$ & FPS $\uparrow$  & Train error $\downarrow$ & Test error $\downarrow$ \\ \midrule
  1       & 160  & 29 & 1.2M     & 13793 & 5.44e-2    & 6.38e-2   \\ \midrule
  2       & 80   & 17 & 2.3M     & 6865 & 4.20e-2    & 5.13e-2   \\ \midrule
  4       & 40   & 11 & 1.9M     & 5056 & 4.70e-2    & 5.38e-2   \\ \midrule
  8       & 20   & 9  & 3.7M     & 2603 & 8.14e-2    & 8.71e-2   \\ \bottomrule
\end{tabular}
\label{tab:partition}
}
\end{minipage}
\end{wrapfigure}

We found that for generating 160 frames in total, bi-partition strikes the best balance between accuracy and parameter efficiency.
Too many partitions indirectly increase the number of iterations, which causes severe distribution shifts and error accumulation.
More ablation studies can be found in Appendix~\ref{sec:HyperparameterPicking}.

\noindent\textbf{Cross-stage model sharing and hidden state propagation}.~ Table~\ref{tab:PartitionTech} compares different methods of partition, which suggests that the best partition strategy is to pass history information between adjacent independent submodels from $\Psi_{\theta_{i-1}}$ to $\Psi_{\theta{i}}$ in addition to the prediction result. 
Simply iterating twice, CNN will suffer from a severe distribution shift. 
By jointly optimizing two weight-sharing CNNs, the model is aware of the error accumulation, thus showing better roll-out errors. 
Once we relax them to two independent CNNs, combined with the hidden state that passes more information to the downstream submodels, we achieve the lowest error in 160 frames of prediction.

\begin{table}[htp]
\centering
\caption{Stage-dedicated models with hidden state propagation give the best 160-frame  fidelity.}
\resizebox{0.92\columnwidth}{!}{
\begin{tabular}{cccc|cc}
\toprule
                           & Hidden state  & Out frames & \#Iter & Single test error & Roll-out test error \\ \midrule
A single CNN               & N/A    & 80         & 2      & 2.13e-2    & 7.30e-2            \\ \midrule
Two weight-sharing CNNs & $\times$ & 80+80        & 1      & 5.98e-2    & 5.98e-2            \\ \midrule
Two independent CNNs  & $\checkmark$ & 80+80        & 1      & 4.86e-2    & 4.86e-2            \\ \bottomrule
\end{tabular}
\label{tab:PartitionTech}
}
\end{table}

\noindent\textbf{Key components in \name architecture}.~
Table~\ref{tab:AblationStudy} shows the performance of different network settings to predict 80 frames of the light field, where we progressively transform to our proposed \name. 
Starting from a simple CNN with the lifting layer~\cite{li2021fourier} in FNO\cite{li2021fourier} as the encoder, first, we replaced it with our convolution encoder, and the test error decreased by 29\%.
When we adopt a dilation factor of 4 with an extra depth-wise convolution ahead to aggregate local information, the error only increased by 2.3\%, but we have 10 times fewer parameters. 
Then, we partitioned the model into two independent submodels, each with a half kernel size, leading to a 17.4\% lower prediction error. 
We then propagate an extra hidden state to the second network for error compensation, which further boosts the fidelity by 6.1\%. 
Finally, we replace the dilated CNN with our DPAConv to introduce permittivity awareness and obtain the best performance with only 1M parameters.

\begin{table}[htp]
\centering
\caption{Ablation study on \name framework. 
Starting from a SimpleCNN with a Lifting field encoder and Conv2d backbone, we progressively add/modify one component.}
\resizebox{\columnwidth}{!}{
\begin{tabular}{l|cccc}
\toprule
                                              & \#Params $\downarrow$ & FPS $\uparrow$ & Train error $\downarrow$ & Test error $\downarrow$\\ \midrule
Baseline SimpleCNN (Lifting Encoder+Conv Backbone)                           & 12M      & 3293    & 2.52e-2    & 3.00e-2   \\ \midrule
~~+Convolutional Field Encoder                         & 12M      & 3235    & 1.74e-2    & 2.13e-2   \\ \midrule
~~~~+Dilated Conv Backbone                  & 1.1M    & 7058    & 1.79e-2    & 2.18e-2   \\ \midrule
~~~~~~+Bi-partition with Stage-dedicated Models ($\Psi_{\theta_1},\Psi_{\theta_2}$) & 0.9M     & 5125    & 1.55e-2    & 1.80e-2   \\ \midrule
~~~~~~~~+Cross-stage Hidden State Propagation   & 0.9M     & 5065    & 1.52e-2    & 1.69e-2   \\ \midrule
~~~~~~~~~~+Device Encoder+DPAConv Backbone (Final settings) & 1M       & 4560    & 1.43e-2    & 1.59e-2   \\ \bottomrule
\end{tabular}
\label{tab:AblationStudy}
}
\end{table}

\vspace{-5pt}
\section{Conclusion and Limitation}
\label{sec:Conclusion}
In this work, we present a physics-inspired causality-aware AI-accelerated FDTD solving framework \name for ultra-fast photonic device simulation.
We deeply analyze and organically incorporate physical constraints into our model primitive designs, honoring space-time causality and permittivity-dependent wave propagation principles.
Cross-iteration error mitigation techniques have been proposed to compensate for the distribution shift issue during time-bundled autoregressive prediction with balanced scalability, long-term prediction fidelity, and efficiency.
Compared to SoTA Fourier-based and convolutional neural operators on three challenging photonic device types, our \name outperforms them with 49.1\% less prediction error and 23.5 times fewer parameters.
300-600 $\times$ speedup has been demonstrated over open-source FDTD solvers on average.
One potential limitation is that our framework still observes rapidly accumulated errors with large-iteration rollout even though we have various error suppression methods.
As a future direction, error suppression during auto-regressive prediction or even a completely new formulation beyond auto-regression will be further investigated to support long timespan optical FDTD simulation.

\newpage

{\small
\bibliographystyle{plain}
\bibliography{./ref/Top_sim,./ref/NN,./ref/NP,./ref/ALG, ./ref/addition}
}
\newpage
\appendix

\section{Appendix}

\subsection{Optical simulation detail}
\label{sec:update_rule}
For $TM^z$ polarized electric field, the FDTD updating rule is shown as follows:
\begin{equation}
\small
\label{eq:FDTDFormula}
\begin{aligned}
H_x^{q+\frac{1}{2}}\left[m, n+\frac{1}{2}\right]&= H_x^{q-\frac{1}{2}}\left[m, n+\frac{1}{2}\right]- \frac{\Delta_t}{\mu[m, n+\frac{1}{2}] \Delta_y}\left(E_z^q[m, n+1]-E_z^q[m, n]\right)\\
H_y^{q+\frac{1}{2}}\left[m+\frac{1}{2}, n\right]&= H_y^{q-\frac{1}{2}}\left[m+\frac{1}{2}, n\right]+ \frac{\Delta_t}{\mu[m+\frac{1}{2}, n] \Delta_x}\left(E_z^q[m+1, n]-E_z^q[m, n]\right)\\
E_z^{q+1}[m, n]= E_z^q[m, n] & +\frac{\Delta_t}{\epsilon[m, n] \Delta_x}\left\{H_y^{q+\frac{1}{2}}\left[m+\frac{1}{2}, n\right]-H_y^{q+\frac{1}{2}}\left[m-\frac{1}{2}, n\right]\right\} \\
& -\frac{\Delta_t}{\epsilon[m, n] \Delta_y}\left\{H_x^{q+\frac{1}{2}}\left[m, n+\frac{1}{2}\right]-H_x^{q+\frac{1}{2}}\left[m, n-\frac{1}{2}\right]\right\},
\end{aligned}
\end{equation}
in which the $m$, $n$, $q$ represent discrete counterpart of $x$, $y$, and $t$ in continuous domain. The $\frac{1}{2}$ shown in the index refers to the points at the middle point of edges in Yee's grid.
\subsection{Dataset Generation}
\label{subsec:dataset}
For MMI , we randomly generate 20 devices to train and 5 devices to test, for MRR, due to it is longer than that of MMI and Metaline, we generate 6 devices to train and 5 to evaluate, for Metaline, 32 device to train and 8 device to test, based on the variable settings and distributions in Table~\ref{tab:AppendixDataset}.
Each device has an individual simulation for each input port.
MMIs sweep over 3 ports and generate a total of 75 simulation videos;
MRRs only have 1 input port and generate a total of 11 simulation videos (much longer);
Metalines sweep over 3 ports and generate a total of 120 simulation videos.
The time interval between two frames is 1 fs, i.e., $\Delta_t=1$ fs.

\noindent\textbf{How to sample video patches as training/validation/test dataset}.~
First, according to different device type, we select different numbers of devices for train and test as demonstrated above.
During training, for each example, we randomly select one device and one port and slice a video segment.
The video segment has a randomly sampled starting frame index of $i$.
$E_{i;i+T_{in}}$ will be the input fields.
When sampling the starting frame index of $i$, considering the imbalanced temporal distribution of the source, which means that the source only exist within the first 560 frames approximately, we attribute more probability to sample the starting frames before 560 frames. To be specific, for MMI and Metaline, starting frames that contains source have twice of the probability to be sampled and for MRR, since the video is much longer than that of MMI and Metaline, the starting frames have 6 times of probability to be sampled.
Sources $J_{T_{in}:T_{in}+T}$ will be extracted from $E_{T_{in}:T_{in}+T}$ at the input port region.
$E_{T_{in}:T_{in}+T}$ serves as the target fields.
Examples across epochs are totally randomly sampled.

For validation and inference, we uniformly slice the videos with an offset of 16 frames, i.e., $i=0, 16, 32,\cdots$.
We do not resample the starting frames with sources in validation and test.
All video slices have bilinear interpolated to have the same spatial resolution $(\Delta_x=\Delta_y=140 nm)$. 
Since our mini-batch size during training and inference is, no padding is added.

\begin{table}[htp!]
\centering
\caption{MMI, MRR, Metaline device configurations in dataset generation.}
\label{tab:AppendixDataset}
\resizebox{0.99\columnwidth}{!}{%
\begin{tabular}{l|ccc|c}
\toprule
\multirow{2}{*}{Variables} & \multicolumn{3}{c|}{Value/Distribution}     & \multirow{2}{*}{Unit} \\ \cmidrule{2-4}
\multicolumn{1}{c|}{}                           &  Tunable MMI $3\times3$           &  MRR         &  Metaline $3\times3$ & \multicolumn{1}{c}{}                      \\ \midrule 
Length                                          & $\mathcal{U}$(20, 30)          & -            & $\mathcal{U}$(8, 10)                        & $\mu m$                                        \\
Width                                           & $\mathcal{U}($5.5, 7)          & -            & Length                       & $\mu m$                                        \\
Radius                                           & -          & $\mathcal{U}($5, 15)            & -                       & $\mu m$                                        \\
Port Length                                     & 3                               & 1                           & 3                               & $\mu m$                                        \\
Port Width                                      & $\mathcal{U}($0.8, 1.0)                   & $\mathcal{U}$(0.5, 0.8)          & $\mathcal{U}($0.8, 1.0)                     & $\mu m$                                        \\
Taper Length                                    & 2                 & -                    & 2                               & $\mu m$                                        \\
Taper Width                                     & Port Width+0.3                  & -                    & Port Width+0.3                               & $\mu m$                                        \\
Ring Bus Width                                     & -                  & $\mathcal{U}$(0.5, 0.8)                    & -                               & $\mu m$                                        \\
Bus Waveguide Gaps                                     & -                  & $\mathcal{U}$(0.1, 0.15)                    & -                               & $\mu m$                                        \\
(\#slots, spacing, $w_{slot}$, $h_{slot}$)                                     & -                  & -                    & (1.4*Length, Length/3, $\mathcal{U}$(0.1,1), $\mathcal{U}$(0.2,0.25))                               & $\mu m$                                        \\
Border Width                 & 1                    & 1                 & 1                            & $\mu m$                                        \\
PML Width                                       & 2                  & 2                  & 2                             & $\mu m$                                        \\
Wavelength range                            & [1.4, 1.65]           & [1.4, 1.65]       & [1.4, 1.65]                  & $\mu m$                                        \\
Permittivity ($\epsilon_{\text{cladding}},\eps_r$)                  & \{2.07, 12.11\}                  & \{1, 6\}        & \{2.07, 12.11\}                          & -                                         \\ 
Video frames                 & 833                  & [800*Radius/3]        & 600                          &  fs                                       \\ \bottomrule
\end{tabular}
}
\end{table}

\subsection{Hyper-parameter Selection}
\label{sec:HyperparameterPicking}

\noindent\textbf{How to determine kernel size $K$}.~Kernel size $K$ is the key design parameter for DPAConv that determines the receptive fields of the model.%
\begin{wrapfigure}[9]{rH}{0.33\textwidth}
\begin{minipage}{0.33\textwidth}
    \centering
    \vspace{-10pt}
    \includegraphics[width=\columnwidth]{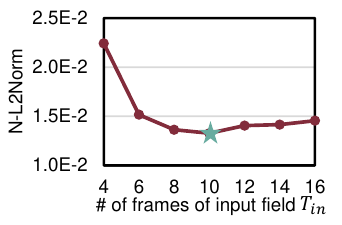}
    \caption{Input fields with $T_{in}=10$ gives the lowest prediction error on MMI.}
    \label{fig:InFrames}
\end{minipage}
\end{wrapfigure}

\begin{figure}
    \centering
    \includegraphics[width=\columnwidth]{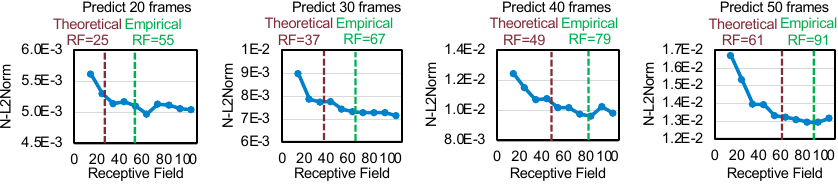}
    \caption{Prediction error (normalized L2-Norm) with different receptive fields (RF) given various output frames.
    A 5-layer CNN with various kernel sizes is trained on the MMI dataset.
    Empirically, using an $\sim 30+$ larger RF than the theoretical one gives the best fidelity and efficiency.}
    \label{fig:CompareReceptiveField}
\end{figure}

To predict the fields after $T$ timesteps, aware of space-time causality, we estimate the furthest distance the wave can propagate in the medium as $R= \sum_0^{T}\frac{c_0\times t}{\sqrt{\epsilon_r}}$. 
Since most light fields are confined in the waveguide region, we use the relative permittivity of the waveguide $\epsilon_r^{wg}$ to calculate the theoretical receptive field $R \approx \frac{Tc_0}{\sqrt{\epsilon_r^{wg}}}$.
Empirically, we recommend a 30 pixels larger receptive field than the theoretical value $R \approx \frac{Tc_0}{\sqrt{\epsilon_r^{wg}}}+30$ to obtain the best fidelity as shown in Figure~\ref{fig:CompareReceptiveField} in which we sweep the receptive field for different timesteps to predict.
Then, each PAConv is assigned to have a receptive field of $[(R-5)/L]$, where 5 is the RF of the field encoder. 
With a dilation factor of $s$, the kernel size of the DPAConv is set to $K=[\frac{R-5}{sL}]$, %
\begin{wrapfigure}[12]{rH}{0.22\textwidth}
    \centering
    \vspace{-28pt}
    \includegraphics[width=0.25\columnwidth]{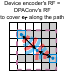}
    \caption{Permittivity along the optical path needed to be encoded.}
    \label{fig:DeviceEncoding}
\end{wrapfigure}
and the kernel size of the precedent depthwise convolution is set to $s+1$.
The device encoder should extract features of the permittivity map along the optical path toward the center pixel, as illustrated in Figure.~\ref{fig:DeviceEncoding}.
Hence, we set the receptive field of the device encoder to $K$.

\noindent\textbf{Frames of input light field $E_{in}$}.~
\name takes the light fields $E_{in}$ from previous timesteps as the initial condition for field prediction.
Given the time locality that we analyzed in Section~\ref{sec:Formulation}, theoretically, two frames ($T_{in}=2$) should provide sufficient information to obtain the current field distribution to calculate the time derivative $\partial E/\partial t$ using first-order finite difference in the Maxwell equation, which indicates the light propagation direction.
Fig.~\ref{fig:InFrames} investigates the impacts of input frames on the prediction error.
We find out that a small number of frames fail to provide enough information for the neural operator to capture the effective initial condition.
At least 8-10 frames are required for the model to deliver low prediction errors.
Note that more timesteps in the input light fields are harmful since the provided information from the further past is irrelevant and useless due to time locality.

\noindent\textbf{Multi-stage partitioning in time-bundling}.~ 
Time bundling is preferred to reduce the iteration times%
\begin{wrapfigure}[13]{rH}{0.43\textwidth}
\begin{minipage}{0.43\textwidth}
    \centering
    \vspace{-10pt}
    \includegraphics[width=\columnwidth]{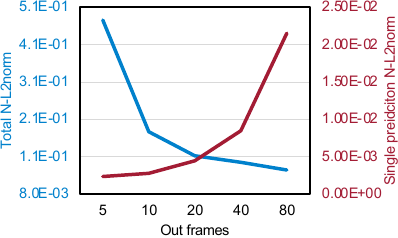}
    \caption{The total roll-out error dominates one single prediction}
    \label{fig:Outframes}
\end{minipage}
\end{wrapfigure}
during the auto-regression. As shown before, there is a trade-off between the speed and memory for different output frames since the output frames are directly related to the kernel size. 
In addition, another trade-off is between the single prediction accuracy and roll-out prediction accuracy. 
Fig~\ref{fig:Outframes} shows the average normalized L2-Norm of the entire 160 frames prediction. 
Fewer output frames make it an easier task and, hence, a smaller single prediction error. 
However, the fidelity benefit obtained from a smaller prediction frame count vanished quickly. For the 160-frame prediction task, the 80 output frames model achieves the best rollout error.

\begin{table}[]
\centering
\caption{Compare different dilation factors $s$ on the Conv2d layers in a SimpleCNN and the MMI dataset in terms of parameter count, runtime, and prediction error (N-L2Norm).
$K_{\text{DWConv}}$, $K_{\text{Conv}}$ represents kernel size for depthwise convolution and Conv2d.
The receptive field roughly remains the same (17$\sim$23).
We select $s$=4 to balance efficiency and fidelity.}
\resizebox{0.8\columnwidth}{!}{
\begin{tabular}{ccc|cccc}
\toprule
    $K_{\text{DWConv}}$ & $K_{\text{Conv}}$ & $s$ & \#Params $\downarrow$ &  FPS $\uparrow$ & Train error $\downarrow$ & Test error $\downarrow$ \\ \midrule
 N/A       & 17 & 1        & 12M      & 3235            & 1.74e-2    & 2.13e-2   \\ \midrule
 3         & 9  & 2        & 3.5M     & 3415            & 1.65e-2    & 2.04e-2   \\ \midrule
 5         & 5  & 4        & 1.1M     & 7058            & 1.79e-2    & 2.18e-2   \\ \midrule
 9         & 3  & 8        & 0.5M     & 9417            & 1.91e-2    & 2.29e-2   \\ \bottomrule
\end{tabular}
\label{tab:dilation}
}
\end{table}

\noindent\textbf{Efficiency-fidelity trade-off in convolution dilation factor}.~

As convolution kernel size increases, especially for dynamic PAConv, the computation and memory cost increase quadratically.
We explore the efficiency-fidelity trade-off with different dilation factors $s$ in Table~\ref{tab:dilation}.
Dilated convolution could speed up the training process dramatically, and combined with a depth-wise local information aggregation convolution, it also achieves high fidelity. Table\ref{tab:dilation} shows different dilation strategies with the same equivalent receptive field. 
Considering the trade-off between fidelity, speed, and parameter efficiency, we choose 4 as the dilation rate.

\noindent\textbf{Light field normalization}.~To increase the generalization and convergence of the model, irrelevant information related to field intensity (light brightness) needs to be normalized.
Besides, proper%
\begin{wrapfigure}[13]{rH}{0.33\textwidth}
\begin{minipage}{0.33\textwidth}
    \centering
    \vspace{-0pt}
    \includegraphics[width=\columnwidth]{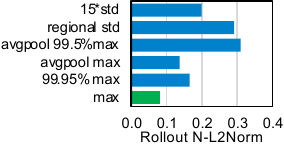}
    \caption{Use max absolute value to normalize and restore the prediction gives us best N-L2norm in one step roll out on MMI.}
    \label{fig:Norm}
\end{minipage}
\end{wrapfigure}
normalization is crucial to minimize error accumulation in autoregressive prediction.
A common method used in PDE learning tasks is to normalize the input field to standard normal distribution per channel and de-normalize the model output using the calculated statistics.
However, this method does not apply to optical FDTD problems with a Gaussian light source where the resultant light field distribution has \emph{high spatial and temporal sparsity}. 
We also observe huge intensity discrepancies across space, time, and data samples.
We evaluate different normalization methods on a 160-frame roll-out test in Fig.~\ref{fig:Norm}. Subtraction of the mean value is not involved in all the normalization method since the mean of the wave is already zero.
To be specific, the first one, 15*std, means that instead of using one standard deviation, we use 15 times of standard deviation to normalize the field since the field are sparse in the devices especially in MRR. 
Use one standard deviation will make the range to be too large.
Second, in regional standard deviation, we first determine the wave propagating region by the average energy and then only calculate the standard deviation within the region for normalization.
The avgpool max refers to the method where we first pick the frame among all the input frames that contains the maximum electric field values. And then reduce the spatial dimension by 8 times. Finally pick the maximum value in the reduced frame.
The avgpool 99.95\% max is almost the same as the avgpool max except that instead of pick the maximum electric field value in the reduced frame, in this method, we pick the 99.5\% quantile maximum value to leave some margin for outliers.
The method called max is straightforward, just pick the maximum absolute electric field value in the input frames to normalize.
The 99.95\% max refers to the method that is almost the same with the former one except instead of using the maximum absolute value to normalize, the 99.95\% employs the 99.95\% quantile to leave some margin for outliers.
Statistics based on standard deviation or 99.95\% quantile show high roll-out error.
Normalizing by the maximum absolute field intensity (\textbf{max}) gives the best roll-out fidelity.

\subsection{Training/inference Settings}
\label{sec:AppendixTrainingSettings}
We adopt Adam optimizer with an initial learning rate of 2e-3, following a cosine learning rate decay schedule and a minimum learning rate of 1e-5 for all the baselines except for the SineNet for which we use the suggested initial learning rate 2e-4 by its author and ended at 1e-6
All models are trained and evaluated on two servers with 8 NVIDIA A6000 GPUs.
The runtime for all neural network models is averaged across 5 runs per photonic device in the test dataset.
The runtime for the CPU numerical FDTD solver MEEP is evaluated on a 64-core AMD EPYC 7763 64-Core Processor.

\subsection{Model Architecture Details}
\label{sec:AppendixModelDetails}
We compare our method with SoTA Fourier-domain neural operators and CNN models.
We set a full mode for Fourier-domain neural operators to enable them to learn local window operations.
For CNN models, we maintain similar layers and the number of parameters for fair comparison. And for Fourier kernel integral operation models, we try to choose as many modes as possible to capture the local wave behavior.

\noindent\textbf{FNO~\cite{li2021fourier}}: We construct a 4-layer FNO with Fourier modes of (128, 128), hidden channel of 36. The total parameters are 340M for FNO and MRR. For Metaline, the data are padded to 168, so the full mode is (68, 68), with hidden channel of 36, the total parameters are 146M.

\noindent\textbf{F-FNO~\cite{tran2023factorized}}: We construct a 12-layer F-FNO whose modes, for MMI and MRR that padded to $256\times 256$ is (128, 129) and for metaline which is padded to $168\times168$, is (84, 85). The number of parameters is 4.5M and 3.3M, respectively.

\noindent\textbf{KNO~\cite{xiong2023koopman}}: We construct a 2-layer KNO whose modes, for MMI and MRR that padded to $256\times 256$ is (128, 129) and for Metaline which is padded to $168\times168$, is (84, 85). The number of parameters is 171.8M and 74.6M, respectively.

\noindent\textbf{NeurOLight~\cite{gu2022neurolight}}: We construct a 6-layer NeurOLight whose modes, for MMI and MRR that padded to $256\times 256$ is (128, 129) and for Metaline which is padded to $168\times168$, is (84, 85). The number of parameters is 2.2M and 1.6M, respectively.

\noindent\textbf{SimpleCNN}: We construct a 16-layer SimpleCNN. 
The kernel size is set to be 15 for MMI, and Metaline has the same $\epsilon_r$, and hence the same required receptive field, and the kernel is set to be 21 for MRR due to the high-speed light because of the relatively small $\epsilon_r$. The number of channels is set to 32 so that the number of parameters is within a reasonable range, which is 3.8M for MMI and Metaline and 7.3M, respectively.

\noindent\textbf{SineNet~\cite{zhang2024sinenet}}: We construct a SineNet with 8 waves for MMI and MRR whose was padded to $256\times256$, the number of downsampling and upsampling blocks in each wave is 4, and the initial hidden channel is 24 so that the number of parameters keeps reasonable.
For Metaline, to cooperate with its size which is $168\times168$, we changed the number of downsampling and upsampling blocks to 3 in each block and to compensate, we set the initial channels to 42 and the number of parameters is 30M

\noindent\textbf{\name}: For the device encoder, we use a single convolution layer by a depth-wise convolution layer followed by layer normalization in ConvNeXt\cite{liu2022convnet} style and GELU, a skip connected is added connecting from input of the depth-wise convolution to the end of GELU. Then, the above structure is copied once and cascaded together to form our device encoder. The output channel for the four convolutional layers is $1\rightarrow72\rightarrow72\rightarrow48\rightarrow48$. 
For MMI and Metaline, the kernel size is $[3~3~5~5]$, and for MMR, the kernel size is $[5~5~5~5]$. The device encoder is padded with replicate mode.
For the field encoder, we use almost the same configuration as the device encoder except for the number of channels; the channels are now becoming $\text{\# of input fields}+\text{\# of sources}\rightarrow72\rightarrow72\rightarrow72\rightarrow72$. 
The kernel size becomes $[1~3~1~3]$, and the fields are padded using zero padding.
For the hidden state adaptor, we use a single point-wise convolution, with input channels of 72+72=144 and output channels of 72.
For the backbone, we use 8 layers. 
For MMI and Metaline, each layer has a local aggregation depth-wise convolution layer whose channel number is 72 and kernel size equals 5, a DPAConv layer with kernel size 5 and dilation 4 to provide enough receptive field followed by layer normalization and GELU, all these modules are included within skip connection. For MRR, the basic structure remains the same except for the kernel size for DPAConv, which becomes 7 to provide a wider receptive field.
For the decoder, we use two point-wise convolutional layers in which the first one lifts the $D$-dimension feature to a 512-dimension vector and the second one projects it back to the required output frames, in our case, 80.
The total parameters are 4.4M for MRR and 2.4M for MMI and Metaline.

\subsection{Prediction Result Visualization}
\label{sec:AppendixVisualization}
\begin{figure}
    \centering
    \includegraphics[width=\columnwidth]{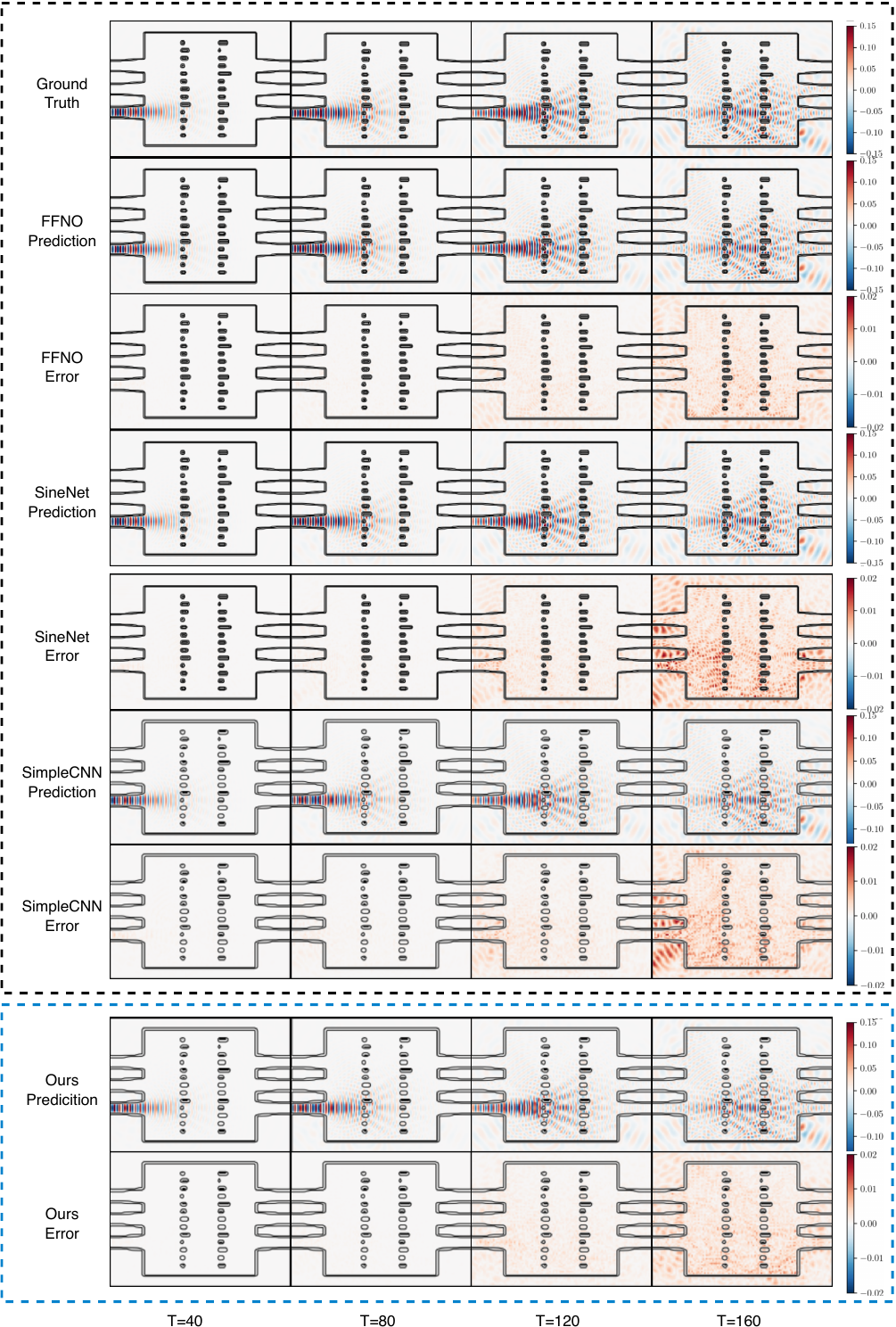}
    \caption{\name Visualization on Metaline, sampled every 40 frames
    }
    \label{fig:VisMeta}
\end{figure}

\begin{figure}
    \centering
    \includegraphics[width=\columnwidth]{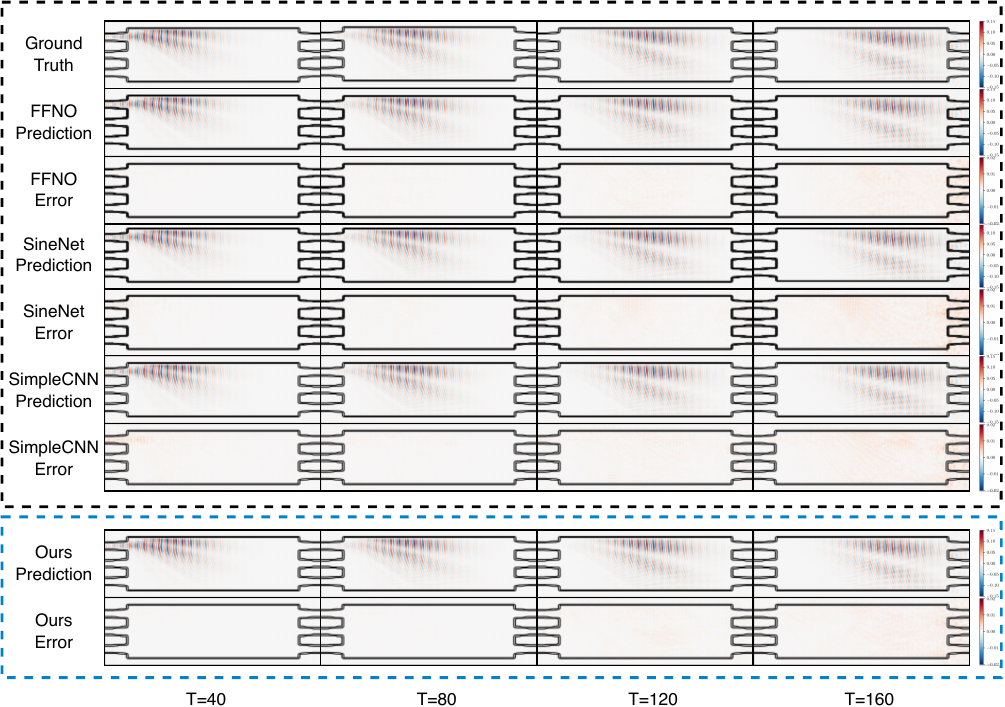}
    \caption{\name Visualization on MMI, sampled every 40 frames
    }
    \label{fig:VisMMI}
\end{figure}

\begin{figure}
    \centering
    \includegraphics[width=0.8\columnwidth]{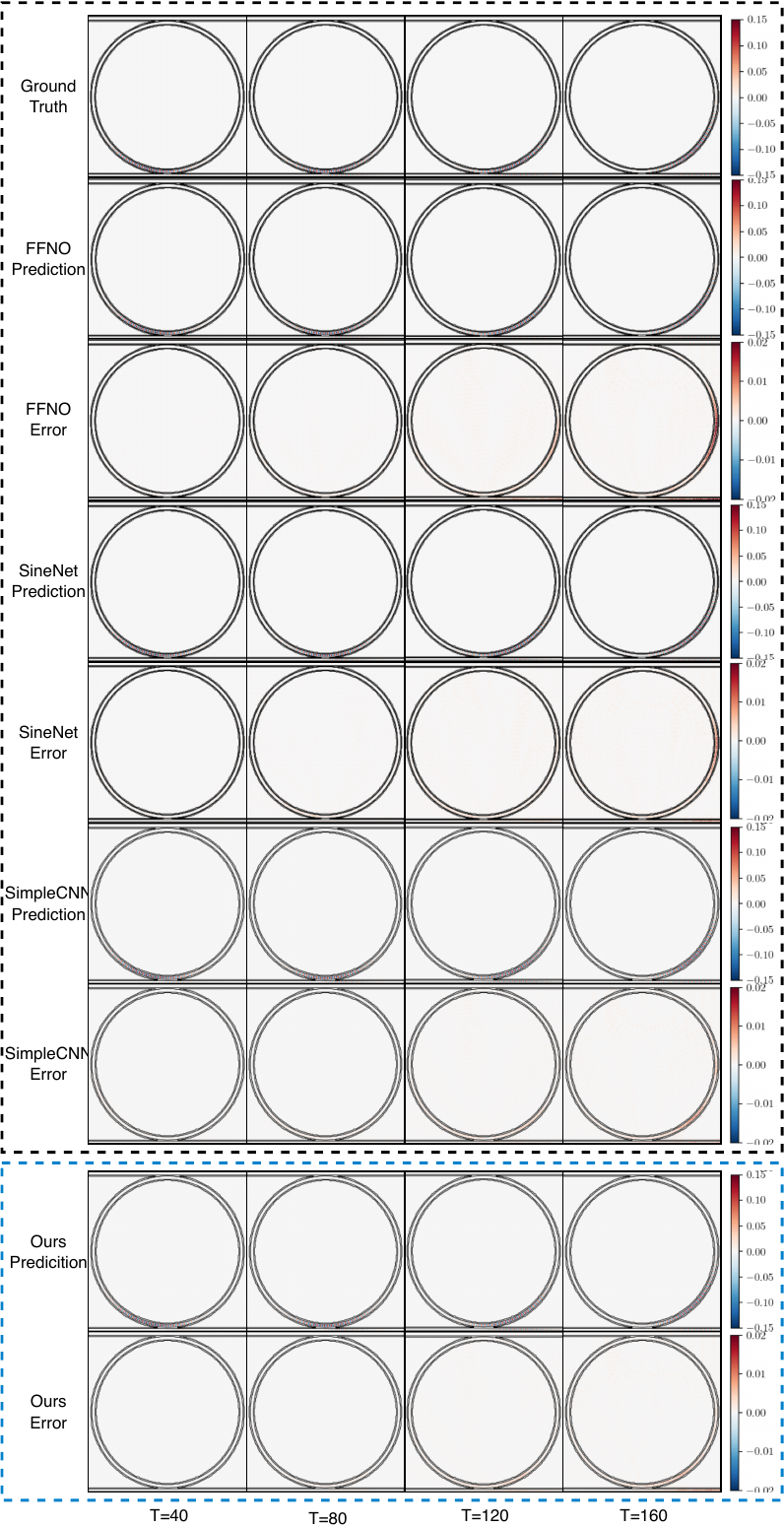}
    \caption{\name Visualization on MRR, sampled every 40 frames
    }
    \label{fig:VisMRR}
\end{figure}

In this section, we select the best performed baselines and our \name to show their performance in 160 frames prediction on different devices. We sampled the 160-frames whole video every 40 frames and to make the error more obvious, the error is plotted with a smaller scale from -0.02 to +0.02

Figure~\ref{fig:VisMMI} shows the visualization of the 160 frames prediction of the selected baselines and \name on MMI. Since the relative simple structure, the error among all the baselines are small. However, there is still an obvious performance gap between \name and other well performed baselines. 

Figure~\ref{fig:VisMRR} shows the visualization of the 160 frames prediction of the selected baselines and \name on MRR in which a pulse of light is propagating through the slim ring waveguide and coupled to the straight waveguide. More complicated device raises a more challenging task for these surrogates. FFNO suffers from the Fourier integral operation and shows huge error. SineNet, also, had a bad performance due to the irrelevant features. CNN based operators obtained better fidelity and the \name, due to its physics causal dynamic kernel, have a slightly better performance than SimpleCNN.

Figure~\ref{fig:VisMeta} shows the visualization of the 160 frames prediction of the selected baselines and \name on Metaline. The Metaline has the most complicated structure, which causes the larger performance gap between \name and SimpleCNN as expected

\end{document}